\documentclass[prl,twocolumn,unsortedaddress,showpacs,floatfix]{revtex4}

\usepackage{graphicx}
\usepackage{array}
\usepackage{amsmath,amsfonts,amssymb}

\begin{document}

  \title{Surface-State Localization at Adatoms}

  \author{L.  Limot}
  \author{E.  Pehlke}
  \author{J.  Kr\"{o}ger}
  \author{R.  Berndt}
  \affiliation{Institut f\"{u}r Experimentelle und Angewandte Physik
  and Institut f\"{u}r Theoretische Physik und Astrophysik, Christian-Albrechts-Universit\"{a}t zu Kiel, D-24098 Kiel, Germany}

  \date{\today}

  \begin{abstract}
    Low-temperature scanning tunneling spectroscopy of magnetic and
    non-magnetic metal atoms on Ag(111) and on Cu(111) surfaces reveals the
    existence of a common electronic resonance at an energy below the binding
    energies of the surface states.  Using an extended
    Newns-Anderson model, we assign this resonance to an adsorbate-induced
    bound state, split off from the bottom of the surface-state band, and
    broadened by the interaction with bulk states. A lineshape analysis of the
    bound state indicates that native adatoms decrease the surface-state lifetime,
    while a cobalt adatom causes no significant change.
  \end{abstract}


  \maketitle

\begin{figure}[t]
    \includegraphics[width=5.5cm,bbllx=65,bblly=285,bburx=530,bbury=755,clip=]{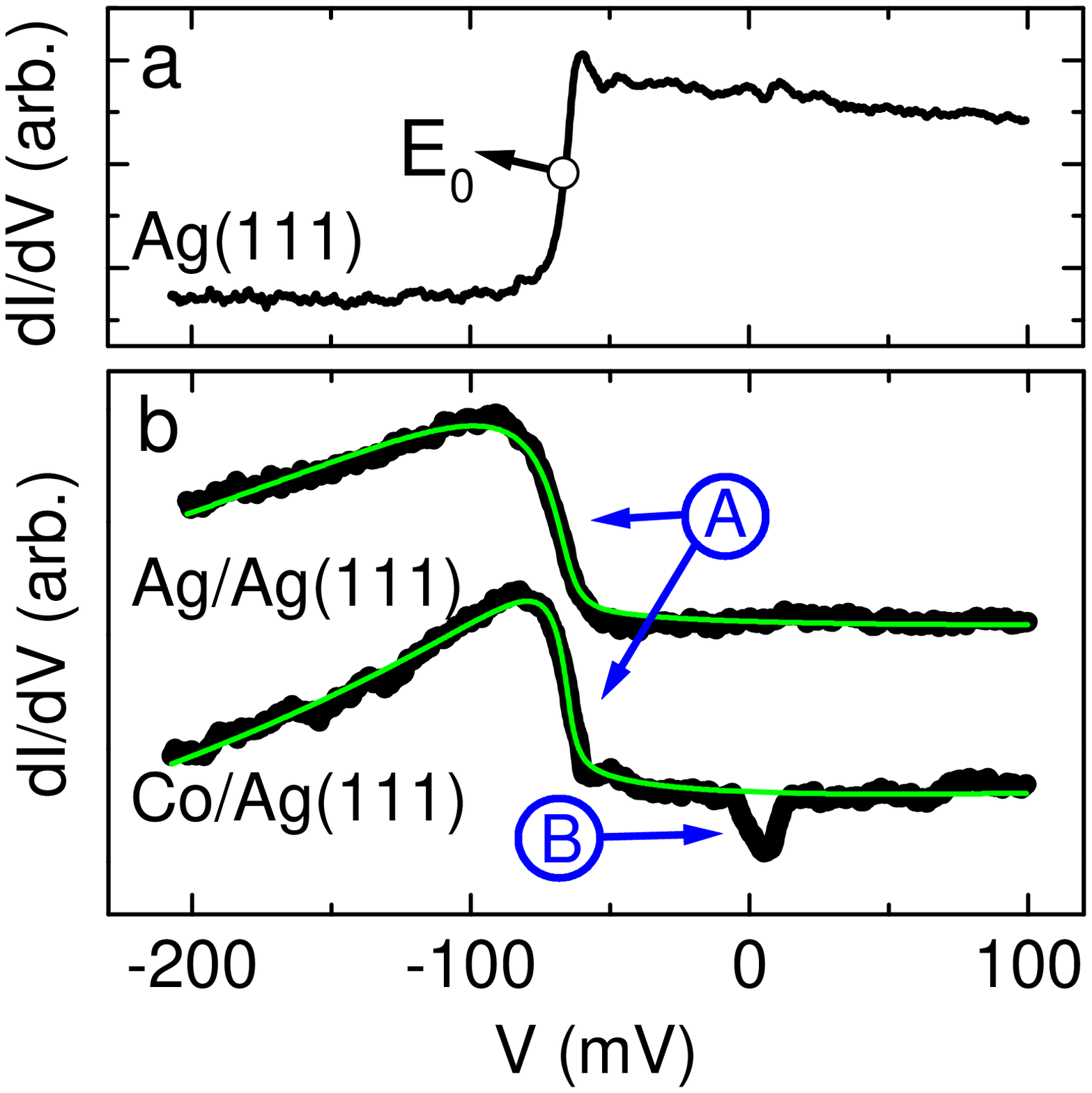}
    \caption{${\rm d}I/{\rm d}V$ spectrum over: (a) Center of a $20\times 20\,{\rm nm}^2$
    defect- and impurity-free area of Ag(111), (b) Center of a silver atom and of a cobalt
    atom on Ag(111) (feedback loop opened at $R=200\,{\rm M}\Omega$, spectra are shifted
    vertically for clarity). Light-colored solid lines: fits described in the text.}
    \label{fig1}
  \end{figure}

\begin{figure}[t]
    \includegraphics[width=5.5cm,bbllx=65,bblly=285,bburx=530,bbury=755,clip=]{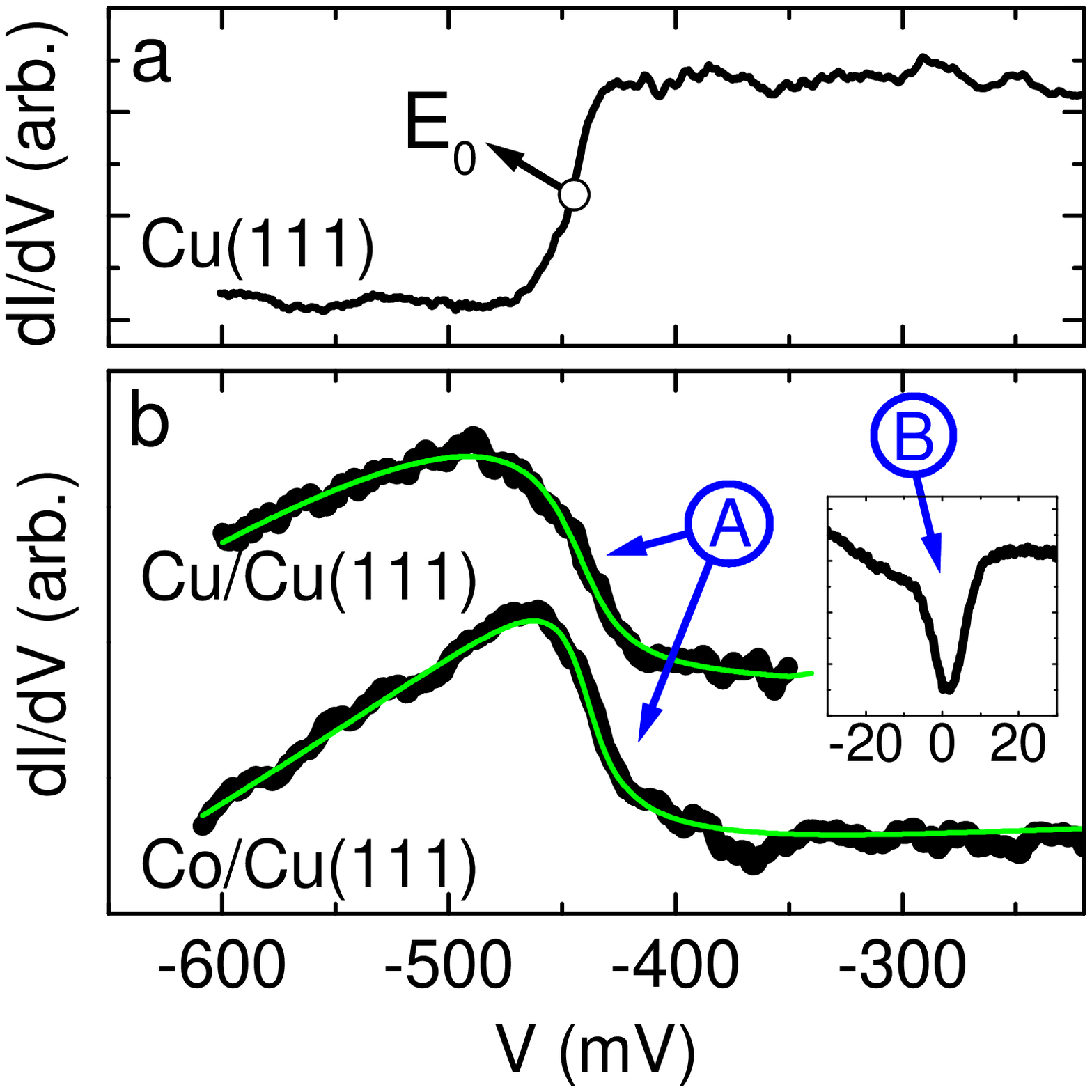}
    \caption{${\rm d}I/{\rm d}V$ spectrum over: (a) Center of a $20\times 20\,{\rm nm}^2$
    defect- and impurity-free area of Cu(111), (b) Center of a copper atom and of a cobalt
    atom on Cu(111) (feedback loop opened at $R=100\,{\rm M}\Omega$, spectra are shifted
    vertically for clarity). Light-colored solid lines: fits described in the text.
    \textit{Inset:} Kondo-Fano resonance of Co/Cu(111) near $E_{\rm F}$.}
    \label{fig2}
  \end{figure}

  The unique ability of scanning tunneling microscopy and spectroscopy (STS)
  to access locally the density of states of single adsorbed atoms and
  clusters has been recently used to investigate magnetic adatoms which -- owing
  to the Kondo effect -- exhibit a sharp spectroscopic structure close to the Fermi
  energy $E_{\rm F}$ \cite{li98,mad98,man00,mad01,kno02,mer04}.
  Surprisingly, only few studies of metal surfaces have been reported for non-magnetic adatoms
  and for a wider energy range around the Fermi level \cite{mad01,avo95,jam00,nil02,kvb04}.
  There is a good reason to explore the physics beyond a narrow range around $E_{\rm F}$.  For instance,
  it is known that a localized attractive perturbation of a two-dimensional
  electron gas should result in the appearance of a bound state, split off from the
  bottom of the continuum, and with a wave function localized around the
  perturbation \cite{eco83}. Surface and image-potential states represent an opportunity
  to investigate this scenario with STS. In fact, it has been predicted that bound states
  should appear around single alkali atoms on metal surfaces as a consequence of their
  attractive perturbation on these two-dimensional electron gases \cite{gau04}.\\
  \- In this Letter, we present a comparative low-temperature STS study of a set
  of adsorbates on the Ag(111) and the Cu(111) surfaces.  We show that the density of states (DOS) of single
  silver and cobalt atoms adsorbed on Ag(111), as well as single copper and cobalt atoms adsorbed
  on Cu(111), exhibit a resonance below the binding energy $E_0$ of the surface states of these substrates.
  Within the framework of a Newns-Anderson model extended to a two-band interaction, we assign this
  resonance to a bound state split-off from the bottom edge of the surface-state band.  A lineshape
  analysis suggests that the scattering of the surface state at the adsorbate affects
  its lifetime, depending on the adatom nature. This appears to be a general property of atoms
  interacting with a two-dimensional electron gas.\\
  \- The measurements were performed in a homebuilt ultrahigh vacuum scanning
  tunneling microscope at a working temperature of $T=4.6\,{\rm K}$. The Ag(111) and
  the Cu(111) surfaces were cleaned by Ar$^{+}$ sputter/anneal cycles. The single copper and
  silver adatoms were created by controlled tip-sample contact, whereas the single
  cobalt atoms were evaporated onto the cold substrates by heating a
  degassed cobalt wire wound around a pure W wire ($>99.95\,\%$).  The
  evaporation, through an opening of the liquid helium shield of the
  cryostat, yielded a coverage of $3\times 10^{-3}\,{\rm ML}$ (ML=monolayer).  No appreciable
  increase of other impurities was detected. The spectra of the differential
  conductance of the tunneling junction, ${\rm d}I/{\rm d}V$ versus $V$, where $V$ is the
  sample bias measured with respect to the tip, were acquired via
  lock-in detection (the root-mean-square (rms) modulation was $1\,{\rm mV}$ in amplitude and
  $\approx 10\,{\rm kHz}$ in frequency) while the current feedback was open.  The etched
  W tip was treated \textit{in situ} by soft indentations
  into the surface, until adatoms were imaged spherically
  and tip-structure artifacts were minimized near the ${\rm d}I/{\rm d}V$-voltages of interest.
  The thermalization of the substrate to $4.6\,{\rm K}$ ensured a negligible thermal diffusion
  of the adatoms within measurement times.\\
  \- Figures~\ref{fig1} and \ref{fig2} illustrate the main experimental
  findings.  A ${\rm d}I/{\rm d}V$ spectrum acquired over a bare terrace of Ag(111) is
  presented in Fig.~\ref{fig1}a.  The DOS has a sharp step-like onset at an
  energy $E_{0}$ which corresponds to the lower edge of the Ag(111)
  surface-state band, with a typical width governed by the surface-state lifetime $\hbar/\Gamma$.
  A similar onset is observed over a terrace of Cu(111) (Fig.~\ref{fig2}a).
  Following Ref.~\onlinecite{kli00}, we extract
  $E_{0}=-67(1)\,{\rm meV}$ and $\Gamma=7(1)\,{\rm meV}$ from the Ag(111) onset, and $E_0=-445(1)\,{\rm meV}$ and
  $\Gamma=24(1)\,{\rm meV}$ from the Cu(111) onset, in agreement
  with recent lifetime measurements by angle-resolved photoemission \cite{rei01}.
  The spectrum acquired over the center of individual
  atoms adsorbed on these surfaces differs substantially from the surface state
  spectrum as shown in Fig.~\ref{fig1}b for a silver adatom (Ag) and a cobalt
  adatom (Co) on Ag(111), and in Fig.~\ref{fig2}b for a copper adatom (Cu) and a cobalt
  adatom on Cu(111). The step-like feature at $E_0$ is lost, and new features appear in the
  DOS of the adsorbates: a resonance below $E_0$ (labeled {\bf A}), and, in addition for
  the cobalt atoms, a resonance near the Fermi energy (labeled {\bf B}).  When the tip is
  moved laterally off the center of the adsorbates, all features continuously
  decrease in amplitude and vanish at $\approx 10\,{\rm\AA}$ away from the
  center.  At a distance of $\agt 50\,{\rm\AA}$, the DOS of the
  surface-state is recovered. When decreasing the tunneling resistance over three orders
  of magnitude, corresponding to a height change between tip and adsorbate of approximately $3\,{\rm\AA}$,
  feature {\bf A} remains unchanged, except for a shift of the spectrum to lower energies,
  similar to the Stark shift observed for surface-state electrons \cite{lim03}.
  The observation of feature {\bf A} and its interpretation are the main topics of this Letter.\\
  The structure near the Fermi energy (feature {\bf B}) on cobalt is a Kondo
  resonance, which is detected as a Fano line in the ${\rm d}I/{\rm d}V$ spectrum
  \cite{li98,mad98,ujs00,mad01,mer04}. Since its first observation,
  several STS studies have reported similar resonances for a variety of Kondo systems
  \cite{man00,mad01,kno02,mer04}. The Fano fit of various cobalt spectra
  yielded Kondo temperatures $T_{\rm K}=83(10)\,{\rm K}$ and $T_{\rm K}=63(6)\,{\rm K}$ for, respectively, Co/Ag(111) and Co/Cu(111),
  in agreement with previous studies on these systems.\\
  \begin{figure}[t]
    \includegraphics[width=7.5cm,bbllx=25,bblly=295,bburx=550,bbury=640,clip=]{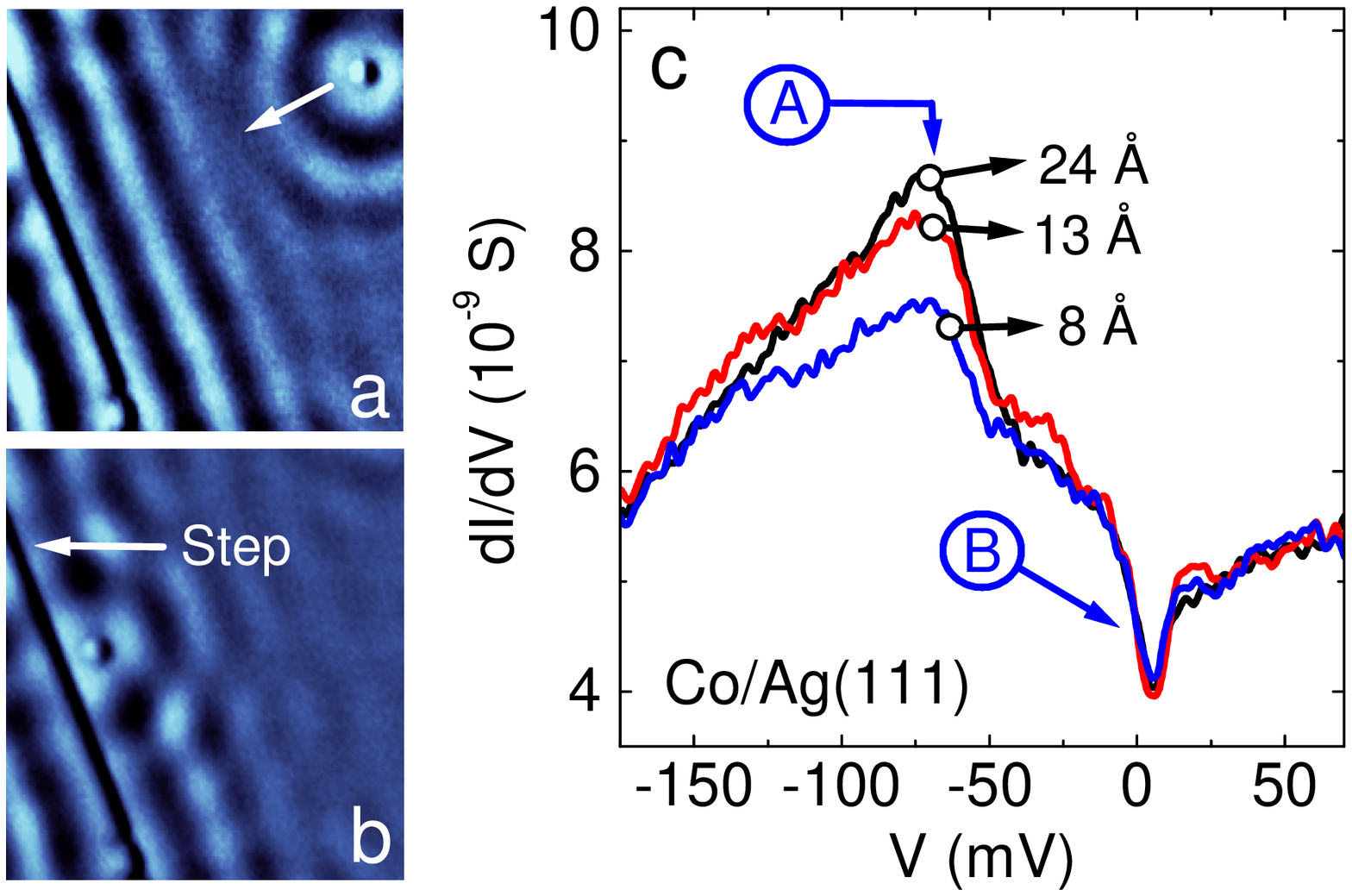}
    \caption{${\rm d}I/{\rm d}V$ images of a cobalt adatom near a monoatomic step of Ag(111)
    ($16\times 16\,{\rm nm}^2$, $I=0.5\,{\rm nA}$, $V=100\,{\rm mV}$).  (a) Prior, (b) After
    manipulation (at $8\,{\rm\AA}$ from the step).  Cobalt atoms are moved at
    $I=50\,{\rm nA}$ and $V=2.8\,{\rm V}$.  (c) ${\rm d}I/{\rm d}V$ spectra of a
    cobalt adatom at $24$, $13$ and $8\,{\rm\AA}$ from the bottom edge of a
    monoatomic step of Ag(111) (feedback loop open at $R=200\,{\rm M}\Omega$).}
    \label{fig3}
  \end{figure}
  The comparison between the Ag/Ag(111) and the Co/Ag(111) spectra, or between
  the Cu/Cu(111) and the Co/Cu(111) spectra, is a clear indication that feature {\bf A}
  is related to a non-magnetic effect contrary to the Kondo effect of the cobalt atoms.
  Since this feature is observed in the same energy range for Co and Ag on Ag(111),
  and for Cu and Co on Cu(111), it is also unlikely that this resonance reflects an atomic orbital of
  the adsorbate, for example a cobalt $d$ state.  It must rather be related to the surface-state
  electrons of these surfaces, since the resonance occurs slightly below $E_0$ regardless of the adsorbate.
  To probe a possible surface-state involvement, additional spectroscopic
  information can be gathered by a close inspection of the adatom DOS near a
  monoatomic step edge of Ag(111) (Fig.~\ref{fig3}).  In this region, the back scattering of the surface-state
  electrons by the step leads to a standing wave pattern, which modifies the surface-state DOS \cite{dav91,has93}.
  In the case of cobalt, the changes in the adatom
  spectra produced by this perturbation can be monitored by combining
  atom-manipulation and spectroscopy.  Figures~\ref{fig3}a and~\ref{fig3}b
  illustrate two steps from a typical manipulation procedure for cobalt
  atoms, and Fig.~\ref{fig3}c presents the spectra acquired over a cobalt
  adatom positioned at distances of $8$, $13$ and $24\,{\rm\AA}$ from the
  step edge.  The overall structure of feature {\bf A} changes in this region.
  A shoulder appears at $-30\,{\rm mV}$ and the amplitude decreases as the cobalt atom
  is approached to the step.  A similar behavior is also observed for
  Ag/Ag(111), suggesting that this resonance has a strong surface state
  contribution.  In contrast, the Kondo-Fano resonance does not reveal any
  appreciable change and $T_{\rm K}$ is constant in this region.  While this may
  hint to a minor role for surface-state electrons in the Kondo effect of Co/Ag(111),
  a recent microscopic two-band theory of the surface-Kondo effect points
  to an intricate scenario \cite{lin04}.\\
  \- From our experimental findings, we conclude that the resonance in the
  DOS of the adsorbates is surface-state-related.  This appears to be a
  general property of noble-metal surface states, as a similar feature is also
  observed in STS of single atoms adsorbed on Au(111) \cite{mad01}.  Along the lines
  of Ref.~\onlinecite{mad01}, the most likely explanation for this resonance is
  that it is an adatom-induced-bound state, split off from the bottom of the
  surface-state band.\\
  \begin{figure}[t]
    \includegraphics[width=5.5cm,bbllx=45,bblly=345,bburx=530,bbury=755,clip=]{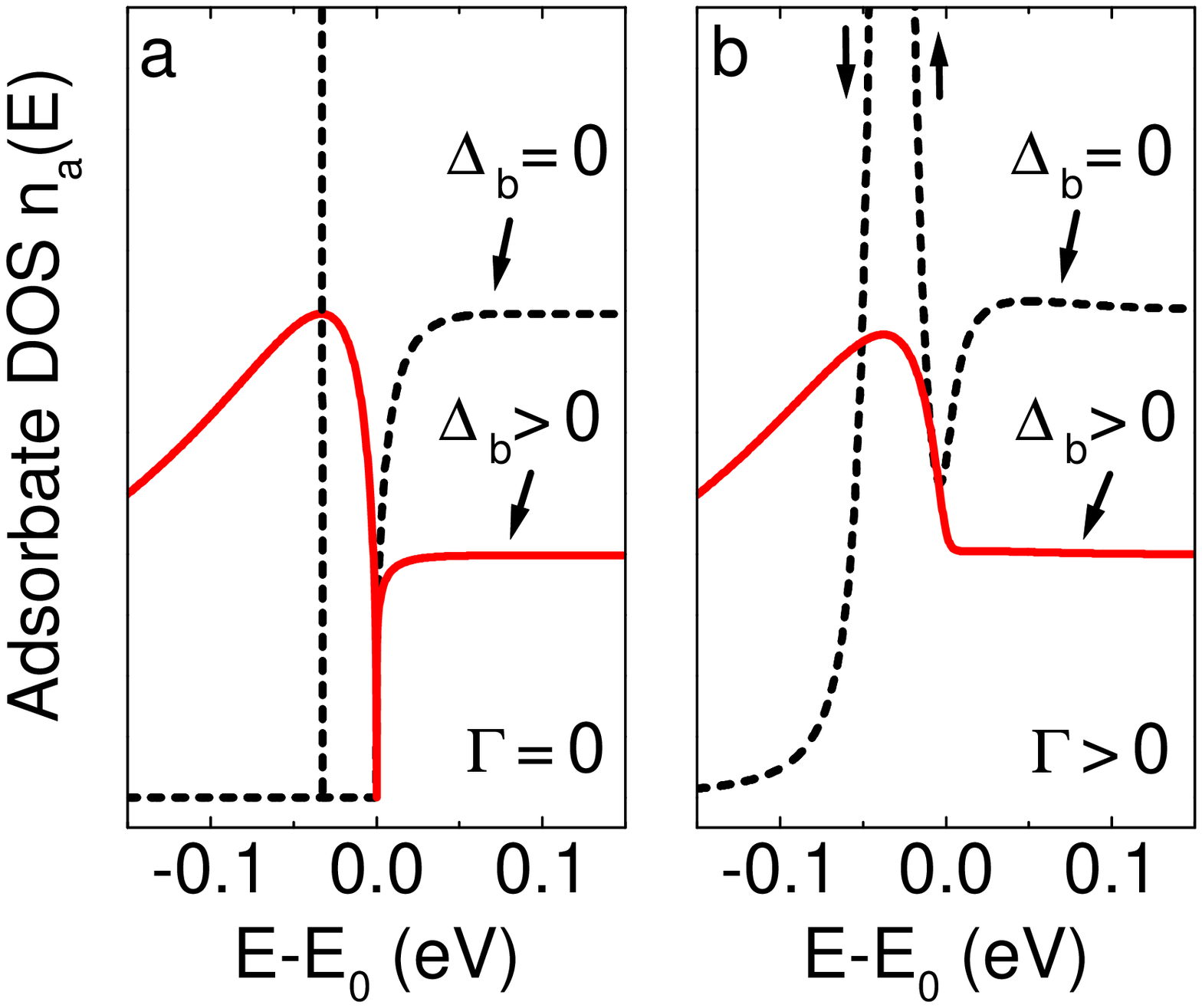}
    \caption{Adatom DOS with a surface-state DOS (a) $n(E)=\Theta(E-E_0)$, (b) including
    lifetime effects. Dashed lines: without a coupling to the bulk states, solid lines: with a
    coupling to the bulk states.  For clarity, part of the bound state is not shown in panel b.}
    \label{fig4}
  \end{figure}
  To model our data we start from a Newns Anderson
  Hamiltonian describing a single adsorbate level $|a\rangle$
  of energy $\varepsilon _a$ interacting with the bulk Bloch states $|{\bf q}\rangle$ \cite{new69},
  and add a second interaction to the surface states $|{\bf k}\rangle$ of the
  (111) substrate:
  \begin{displaymath}
    \mathcal{H} =
      \begin{pmatrix}
        \varepsilon_a             & V_{a\mathbf{q}}          & \tilde{V}_{a\mathbf{k}}          \\
        \tilde{V}_{a\mathbf{q}}^* & \varepsilon_{\mathbf{q}} & 0                                \\
        \tilde{V}_{a\mathbf{k}}^* & 0                        & \tilde{\varepsilon}_{\mathbf{k}}
      \end{pmatrix},
  \end{displaymath}
  where $\varepsilon_{\mathbf{q}}$ and $\tilde{\varepsilon}_{\mathbf{k}}$ denote submatrices
  containing the eigenenergies of the bulk and the surface states, respectively, while $V_{a\mathbf{q}}$
  and $\tilde{V}_{a\mathbf{k}}$ are the coupling matrix elements between the adsorbate level $|a\rangle$
  and the bulk and surface states of the substrate (here $\mathbf{k}$ labels a two-dimensional
  Bloch wave vector, while $\mathbf{q}$ is a tree-dimensional vector).
  The tunneling spectrum with the tip at the
  position of the adsorbate atom is represented, with respect to all
  spectral features, by the projected DOS at the adsorbate
  atom $n_a(E)$.  Applying standard Green's function techniques we
  obtain
  \begin{equation}
    n_a(E) = {1 \over \pi} {\Delta(E) \over \left[E - \varepsilon_a - \Lambda(E)\right]^2 + \Delta(E)^2 } .
    \label{na}
  \end{equation}
  The imaginary part of the self energy $\Lambda(E) + {\rm i} \Delta(E)$
  consists of contributions from the coupling of the adsorbate level to the
  bulk and the surface states, respectively: $\Delta(E) = \Delta_b(E) + \Delta_s(E)$.
  While the bulk contribution
  \begin{equation}
    \Delta_b(E) = \pi \sum_{\mathbf{q}} |V_{a{\mathbf{q}}}|^2\delta(E - \varepsilon_{\mathbf{q}}) = \Delta_b
    \nonumber
  \end{equation}
  is taken to be constant over the energy range of interest, the surface state
  contribution
  \begin{equation}
    \Delta_s(E) = \pi \sum_{\bf k} |\tilde V_{a{\bf k}}|^2 \delta(E - \tilde\varepsilon_{\bf k}) = \Delta_s \Theta(E - E_0)
    \nonumber
  \end{equation}
  is assumed to be governed by the step-function like behavior of the surface state. The real part of the
  self energy follows from $\Delta(E)$ via a Hilbert transform.
  The stepped surface-state DOS results in a logarithmically
  divergent term at $E_0$ as remarked by Borisov {\it et al.} \cite{bor99},
  \begin{equation}
    \Lambda(E)={\Delta_s \over \pi} \ln |E-E_0| + \text{const.},
    \nonumber
  \end{equation}
  where the constant term is absorbed in a renormalized eigenenergy $\varepsilon'_a$ of the adsorbate level.
  The DOS at the adsorbate $n_a(E)$ is presented in Fig.~\ref{fig4}a.  Considering first
  only the coupling of the adsorbate to the surface state (dashed line in Fig.~\ref{fig4}a),
  the adsorbate-substrate interaction produces at the adatom site in case of $\varepsilon_a > 0$
  a bound state at an energy below, but exponentially close to the surface state continuum.
  The additional coupling to the bulk states broadens then the bound state into a resonance
  (solid line in Fig.~\ref{fig4}a).\\
  \- For a quantitative analysis of our results we need to model the surface state DOS as accurately
  as possible.  This is done by including lifetime effects in the surface-state electrons leading to
  a smoother onset of the band edge at $E_0$ as seen experimentally (Figs.~\ref{fig1}a and~\ref{fig2}a). Following Ref.~\onlinecite{li98b},
  we model the surface-state DOS as $n(E)=1/2+\arctan\left[(E-E_0)/(\Gamma/2)\right]/\pi$,
  $\Gamma$ being the inverse lifetime of the surface state. The real part
  of the self energy then reads $\Lambda(E)=\Delta_{s}\ln\left[(E-E_0)^2+(\Gamma/2)^2\right]/2\pi+ \text{const}$.
  Consequently, the bound state acquires an increased linewidth (Fig.~\ref{fig4}b), and a smoother variation
  near $E_0$ occurs. We have used this surface-state DOS to reproduce the experimental data. Results are
  shown as solid lines in Figs.~\ref{fig1}b and~\ref{fig2}b, and the corresponding fitting parameters are
  collected in Tab.~\ref{tab1}.\\
  \- Given the strong coupling $\Delta_b$ to the bulk conduction band found, the
  adsorbate level $|a\rangle$ is most likely the outermost orbital of the adatoms considered here, a $4s$
  level for Co and Cu, and a $5s$ level for Ag. The $d$ orbitals generally lead to
  resonances $\approx 0.1\,{\rm eV}$ wide in the adatom DOS \cite{jam00}, whereas widths of $\approx 1-2\,{\rm eV}$
  have been predicted for $s$ resonances \cite{lan87}. Density functional calculations
  for Ag/Ag(111) confirm this assignment, as the $d$ states of Ag to lie more than $3\,{\rm eV}$ below $E_0$.
  \cite{mac90}
  As for the coupling to the surface-state conduction band, $\Delta_s$ is weaker than $\Delta_b$ for all
  adatoms, but close to the findings by Heller \textit{et al.} of $\Delta_s\sim\Delta_b$
  for adatoms forming a quantum corral \cite{hel94}.\\
  \- Focusing on $\Gamma$, we find that the inverse lifetime required to model the DOS is larger for atoms adsorbed
  on Cu(111) compared to Ag(111), reflecting the slope change near $E_0$ in the spectra of
  Figs.~\ref{fig1}b and ~\ref{fig2}b. This difference is related to the observed inverse lifetimes
  of the clean surfaces (denoted $\Gamma_0$ in Tab.~\ref{tab1}), which is larger for Cu \cite{kli00}.
  However, while
  $\Gamma$ is nearly identical to $\Gamma_0$ for cobalt adatoms, a $50\,\%$
  increase of $\Gamma$ compared to $\Gamma_0$ is observed for the native adatoms.
  While an increased value of $\Gamma$ is consistent
  with the result of a recent photoemission study of the scattering
  of image-potential states by Cu atoms on Cu(001) \cite{bog04},
  the intriguing difference between Co and native adatoms deserves further investigation.\\
  \- STS of Ag-clusters on Ag(111) (not shown) also reveals the existence of a feature at $E_0$,
  suggesting that a bound state is produced regardless of the type of adsorbate -- atom or molecule.
  In fact, the unique condition for a bound state to exist is that an energy level of the adsorbate must be
  coupled to the surface-state conduction band.  However, its
  hybridization with the surface band structure and any change in the DOS of the surface-state electrons
  -- for example near a step edge of the surface (Fig.~\ref{fig3}) or in a nano-cavity -- will modify
  the spectroscopic signature of the bound state.\\
  \begin{table}
    \caption{Parameters of Eq.~\ref{na} employed to reproduce the experimental data.
    $E_0$ was fixed to the surface-state energy onset (Figs.~\ref{fig1}a and~\ref{fig2}a).
    $\Gamma_0$ is the inverse surface-state lifetime extracted from the surface-state DOS
    of the clean surfaces.}
    \begin{ruledtabular}
      \begin{tabular}{rrrrrr}
               & $\varepsilon_a'$ (eV) & $\Delta_b$ (eV) & $\Delta_s$ (eV) & $\Gamma$ (meV) & $\Gamma_0$ (meV) \\
        \colrule
        Ag/Ag  &  0.5(2)               & 0.8(2)          & 0.5(2)          & 12(1)          & 7(1)  \\
        Co/Ag  &  0.5(1)               & 0.7(1)          & 0.4(1)          & 7(1)           & 7(1)  \\
        Cu/Cu  & -0.2(2)               & 0.9(3)          & 0.3(1)          & 35(5)          & 24(1) \\
        Co/Cu  &  0.1(2)               & 0.9(2)          & 0.3(1)          & 24(2)          & 24(1) \\
      \end{tabular}
    \end{ruledtabular}
    \label{tab1}
  \end{table}
  To summarize, single magnetic and non-magnetic atoms interacting with surface-state electrons
  produce a bound state in a energy range exponentially close to the surface-state low band edge. The bound state
  results from the coupling of the adatom's outermost orbital with the surface-state electrons, and is broadened by
  the interaction with bulk electrons. An analysis of the bound-state lineshape indicates that the surface-state
  lifetime is altered by the presence of the native adatoms whereas Co adatoms do not cause a prominent change.\\
  \- We gratefully acknowledge discussions with J.\ P.\ Gauyacq, and thank the Deutsche Forschungsgemeinschaft for
  financial support.

\end{document}